\documentclass[conference]{IEEEtran}
\IEEEoverridecommandlockouts
\usepackage{cite}
\usepackage{amsmath,amssymb,amsfonts}
\usepackage{algorithmic}
\usepackage{graphicx}
\usepackage{textcomp}
\usepackage{xcolor}
\usepackage{url}
\usepackage{multirow}
\usepackage{subcaption}
\usepackage{hyperref}
\usepackage{adjustbox}
\usepackage{booktabs, multirow, makecell} % for nice tables
\newcolumntype{P}[1]{>{\raggedright\arraybackslash\noindent}p{#1}} % noindent column type for booktabs

\def\BibTeX{{\rm B\kern-.05em{\sc i\kern-.025em b}\kern-.08em
    T\kern-.1667em\lower.7ex\hbox{E}\kern-.125emX}}
    
\usepackage{fancyhdr}
\begin{document}

\title{Does the Heart Show Your Pain? Tackling the X-ITE Pain Challenge with Self-Supervised ECG Representation Learning

\thanks{This work was partially supported by 
Polish National Science Centre, Poland, projects no. 2020/37/B/ST6/03806 (D.K., P.K.) and 2024/53/B/HS6/01256 (S.S.);
the statutory funds of the Department of Artificial Intelligence, Wroclaw University of Science and Technology;
the Polish Ministry of Education and Science within the programme “International Projects Co-Funded”;
the European Union under the Horizon Europe, grant no. 101086321 (OMINO). However, the views and opinions expressed are those of the author(s) only and do not necessarily reflect those of the European Union or the European Research Executive Agency. Neither the European Union nor European Research Executive Agency can be held responsible for them.
}}
 %\author{\IEEEauthorblockN{\textcolor{brown}{*** double blind policy ***} }

\author{\IEEEauthorblockN{Dominika  Kunc\textsuperscript{*}, Przemysław Kazienko, Stanisław Saganowski
\IEEEauthorblockA{
\textit{Department of Artifcial Intelligence, Wrocław University of Science and Technology, Wrocław, Poland} \\
\textsuperscript{*}dominika.kunc@pwr.edu.pl}
}
}
\maketitle
\thispagestyle{fancy}

\begin{abstract}
Accurate recognition of pain using physiological signals remains a challenging problem due to pain's subjective nature and high inter-individual variability. In this study, we investigate self-supervised representation learning (SSL) methods applied to unimodal electrocardiogram (ECG), complemented by multimodal pretraining, including accelerometer (ACC) signals from the chest. We focus on classifying low versus medium pain levels on the X-ITE Pain dataset. Our results reveal that while ECG-based models show limited classification performance, multimodal pretraining improves learned representations by capturing cross-modal dependencies. Notably, we observe substantial inter-subject variability in model performance, suggesting that pain-related ECG patterns may be subject-specific. Visualizations indicate distinct subject-specific clustering but no clear separation by pain levels, highlighting the complexity of pain detection from ECG alone. We discuss limitations of unimodal input, label noise, and generalization across subjects and propose future directions. This work advances the understanding of physiological signal representation learning for pain recognition and sets the stage for more robust, clinically relevant wearable pain monitoring solutions.
\end{abstract}


%\vspace*{-0.5\baselineskip}
%\textcolor{red}{\textbf{Workshop Theme - 
%Database and Annotation Design:} Proposals, discussions, or comparisons of different approaches to data collection and annotation to facilitate the study of ambiguity and subjectivity, e.g., stimulus types, measurement instruments, annotation software, and emotion representations.}

%\textcolor{red}{\textbf{DOUBLE BLIND; Deadline 12.06.2024 -> 19.06.2024}}

\vspace*{-0.5\baselineskip}
\section{Introduction and related work}

Wearable devices are becoming increasingly popular for continuous health monitoring \cite{saganowski2022emotion} and offer a promising avenue for objective pain assessment. Unlike self-reports, which are subjective and not always feasible (e.g., in children or non-communicative patients), physiological signals can provide an ongoing stream of data that reflects autonomic nervous system activity \cite{kunc2023emognition, saganowski2021system, kunc2022real}. Several research studies have explored pain recognition using signals such as electrodermal activity (EDA), photoplethysmography (PPG), respiration (RESP), and heart rate variability (HRV) \cite{fernandez2023multimodal, kasaeyan2021pain, avila2021wearable} or ECG \cite{Leng2024}. Among these, EDA has frequently emerged as the most informative \cite{fernandez2023multimodal}, while HRV features extracted from ECG have shown moderate effectiveness in classifying pain intensities under controlled conditions \cite{kasaeyan2021pain}.

Despite these promising findings, translating such approaches to real-world use remains challenging. Many physiological modalities used in experimental studies — like skin conductance or respiration — are not available in most commercial wearables. Furthermore, datasets like X-ITE Pain \cite{gruss2019multi} include multiple modalities, but collecting such rich data in everyday contexts is often impractical. As a result, there is a strong incentive to explore whether a single, easily accessible modality like ECG can support pain assessment outside the lab.

ECG is widely supported in smartwatches, chest straps, and health patches. It can be monitored relatively unobtrusively and raises fewer privacy concerns than cameras or microphones. However, many confounding factors influence ECG, including stress, physical activity, and individual cardiovascular differences, which may obscure pain-related patterns. Some review articles have noted that while wearable-based pain assessment shows promise, especially for chronic pain, further research is needed to improve its robustness and generalizability \cite{avila2021wearable, fernandez2023systematic}. It also refers to chest pain \cite{Cho2025}.

In this work, we focus on ECG-only pain recognition and evaluate the effectiveness of self-supervised learning (SSL) for representation learning. SSL methods can leverage large volumes of unlabeled physiological data to learn general-purpose features, potentially capturing subtle pain-related dynamics that classical feature engineering might miss \cite{Kunc2025PerComPhD, Kunc2025PerComPhD, Avramidis2024}. We pretrain on unlabeled ECG data and fine-tune on the X-ITE Pain dataset \cite{gruss2019multi} for binary pain classification (low vs medium pain).

Overall, the task remains challenging: average F1 scores across subjects are only slightly above random guessing. However, a more detailed analysis reveals substantial inter-subject variability. The model performs surprisingly well for some individuals, suggesting that ECG may carry informative pain-related patterns in certain cases. These findings highlight the potential of personalized approaches and call for further research into subject-specific modeling and adaptation.

% \subsection{Motivation}
% - X-ITE Pain dataset offered many modalities, however monitoring them in the wild is not feasible for most of them 

% - Thanks to wearable devices we can monitor physiological signals relatively easily in daily life. However not all phsyiological signals are that popular in commercially available devices - e.g. not many devices have Skin Conductance, or even temperature. 

% - Therefore we decided to use only ECG signal. It can be monitored in a relatively unobtrusive way, and does not yield as much privacy concerns and anxiety in users as video or audio monitoring. 

% - Our approach is based on self supervised representation learning (SSL), which incorporates self supervised pretraining on vast amounts of unlabeled physiological data. The main goal is to create a representation that will capture the most informative characteristics of the signal, and thus will generalize well to different downstream classification tasks. 

\section{Methodology}

\subsection{X-ITE Pain dataset}
The X-ITE Pain Database \cite{gruss2019multi} was recorded from 134 participants at Otto von Guericke University Magdeburg, Germany, in collaboration with the  Medical Psychology department of the University Clinic Ulm.
It comprises video of the face (frontal and side-view color video and frontal thermal video), video of the body (color and depth), audio, EDA, ECG, and surface electromyography (sEMG) at the trapezius, corrugator, and zygomaticus muscles recorded during phasic (short) and tonic (1 minute long) pain stimulation by heat and electricity, each with three intensities.
In total, it includes about 24,000 phasic and 800 tonic pain stimuli and reactions \cite{gruss2019multi}.

For this study, we used only the challenge subset of the dataset, which includes 26 subjects for training and four subjects for testing, focusing exclusively on low and medium pain levels. Our approach utilizes only the ECG signal from this dataset.

\subsection{SSL -- Representation Learning Model}

We use a unimodal adaptation of the TS-TCC framework \cite{eldele2021time}, a contrastive self-supervised learning method designed for time-series data. The model is trained on 1-lead ECG recordings to learn general-purpose representations without requiring labels. TS-TCC leverages two key contrastive objectives: temporal contrasting, which aims to predict future representations from past context, and contextual contrasting, which encourages consistency across augmented views of the same signal. In addition to the unimodal setup, we also consider a multimodal adaptation of the framework, where the role of augmentations is replaced by two complementary physiological modalities: 1-lead ECG and chest-mounted accelerometry (ACC). During the multimodal pretraining phase, the model learns cross-modal dependencies between ECG and ACC signals. Importantly, the overall framework enables inference from ECG alone, while retaining the enriched representations learned through multimodal pretraining.

To pretrain the SSL model, we used a real-world dataset comprising continuous ECG recordings from 24 individuals collected over four weeks.
The signals were divided into 10-second segments, cleaned using NeuroKit2 \cite{makowski2021neurokit2}, resampled to 100 Hz, and z-scored (within subject). Minimal preprocessing was applied to preserve the natural characteristics of the signal.

The encoder architecture was extended with two additional convolutional layers compared to the original TS-TCC to capture ECG-specific features better. We applied weak and strong augmentations during training, and training was guided by the temporal and contextual contrastive losses. 

After pretraining, the learned representations were fine-tuned on the X-ITE Pain dataset for the downstream pain classification. As the classification head, we used a simple two-layer MLP with a ReLU activation function between the layers. The models were trained using a leave-one-subject-out (LOSO) cross-validation, so each split contained 23 training subjects, two validation subjects, and one test subject.

\section{Results}

%wyniki ogólne i wyniki szczegółowe per osoba, żeby pokazać wariancję i że dla jednych dobrze działa, a dla innych fatalnie działa

%próba analizy czemu nie działa dla części osób nie jest możliwa bo nie mamy informacji o tych osobach

We evaluated the performance of the SSL-pretrained ECG model on the binary pain classification task (low vs. medium pain) using the X-ITE Pain dataset. We tested two pretraining variants for the SSL encoders (i.e., unimodal and multimodal pretraining, described in Section II B.), and two types of classification training: (1) with frozen encoder (only the weights of the classification head were updated in supervised training), and (2) with fine-tuned encoder (all weights were updated in supervised training). The model was fine-tuned on labeled data from each subject and evaluated in a LOSO setting.

\subsection{Overall Performance}
\begin{figure}[ht]
  \centering
  \includegraphics[width=1\columnwidth]{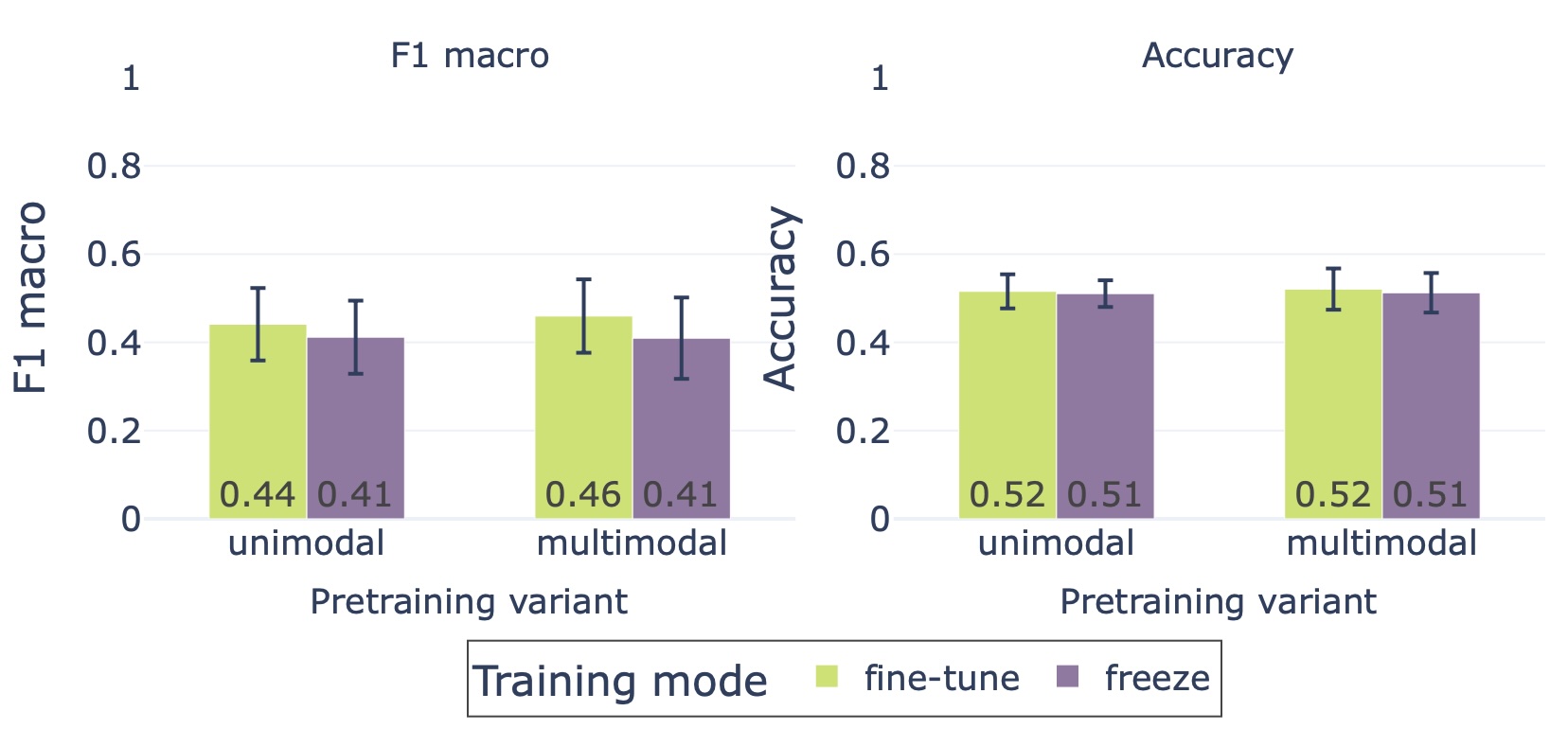}
  \caption{Average (1) F1 macro (2) accuracy values for binary pain classification in the leave-one-subject-out cross-validation.}
  \label{fig:f1_accuracy}
\end{figure}

Figure \ref{fig:f1_accuracy} summarizes the model's overall performance across all subjects. We report the macro-averaged F1-score and classification accuracy. 

In each comparison, we can observe that the fine-tuned models perform better than the frozen ones. Another observation is that the models pretrained in a multimodal setup perform better, indicating the benefits of a cross-modal approach. As the modality in pretraining was ACC collected from the chest, the model could learn some features associated with breathing-related chest movement. This information could enhance pain recognition, given that pain affects respiration by elevating its flow, frequency, and volume \cite{jafari2017pain}.

 The average F1-score from the best-performing model was 0.51, and the average accuracy was 0.54. These values indicate that the model performs only slightly better than random guessing, highlighting the difficulty of the task when relying on ECG data alone.

%Table X: Overall binary classification results (macro F1-score and accuracy).

%Metric	Value
%F1-score	[XX]
%Accuracy [\%]	[XX]\%

\subsection{Subject-Level Performance}
To better understand model behavior, we examined per-subject classification results of the best-performing model. Table \ref{tab:per_participant_results} shows the F1-score and accuracy for each subject. The results reveal substantial inter-subject variability: for some individuals, the model achieved F1-scores exceeding 0.7, while for others, performance dropped close to random levels.

However, we could not investigate why the models performed poorly for specific individuals due to the lack of more detailed data, such as demographic information, specific pain induction indicators, or pain perception scales. Nevertheless, we hypothesize that incorporating personalized information into the models could enhance their performance, and exploring these individual-level differences represents a promising direction for future research.

%Table Y: Per-subject binary classification performance (F1-score and accuracy).

%Subject ID	F1-score	Accuracy [\%]
%S01	[XX]	[XX]\%
%S02	[XX]	[XX]\%
%...	...	...
%SNN	[XX]	[XX]\%

\begin{table}[ht]
\centering
\caption{Per subject performance metrics of the best-resulting model. Subjects in \textbf{bold} performed better than others.}
\begin{tabular}{lcc}
\toprule
\textbf{Subject ID} & \textbf{F1 Macro} & \textbf{Accuracy} \\
\midrule
S008  & 0.511 & 0.550 \\
S009  & 0.549 & 0.563 \\
S012  & 0.491 & 0.500 \\
\textbf{S019}  & \textbf{0.599} & \textbf{0.600} \\
S026  & 0.425 & 0.517 \\
S034  & 0.549 & 0.567 \\
S036  & 0.464 & 0.550 \\
\textbf{S045}  & \textbf{0.705} & \textbf{0.708} \\
S049  & 0.369 & 0.517 \\
S051  & 0.521 & 0.525 \\
S052  & 0.437 & 0.475 \\
\textbf{S057}  & \textbf{0.573} & \textbf{0.592} \\
S068  & 0.335 & 0.504 \\
\textbf{S069}  & \textbf{0.630} & \textbf{0.633} \\
S070  & 0.539 & 0.542 \\
S072  & 0.390 & 0.433 \\
S076  & 0.524 & 0.525 \\
S082  & 0.533 & 0.533 \\
S087  & 0.510 & 0.525 \\
S095  & 0.475 & 0.500 \\
S107  & 0.472 & 0.475 \\
\textbf{S110}  & \textbf{0.602} & \textbf{0.608} \\
\textbf{S114}  & \textbf{0.569}& \textbf{0.583} \\
\textbf{S126}  & \textbf{0.600} & \textbf{0.600} \\
S129  & 0.495 & 0.504 \\
S134  & 0.414 & 0.467 \\
\bottomrule
\end{tabular}
 \label{tab:per_participant_results}
\end{table}

\subsection{Performance Distribution}
\begin{figure}[ht]
  \centering
  \includegraphics[width=\columnwidth]{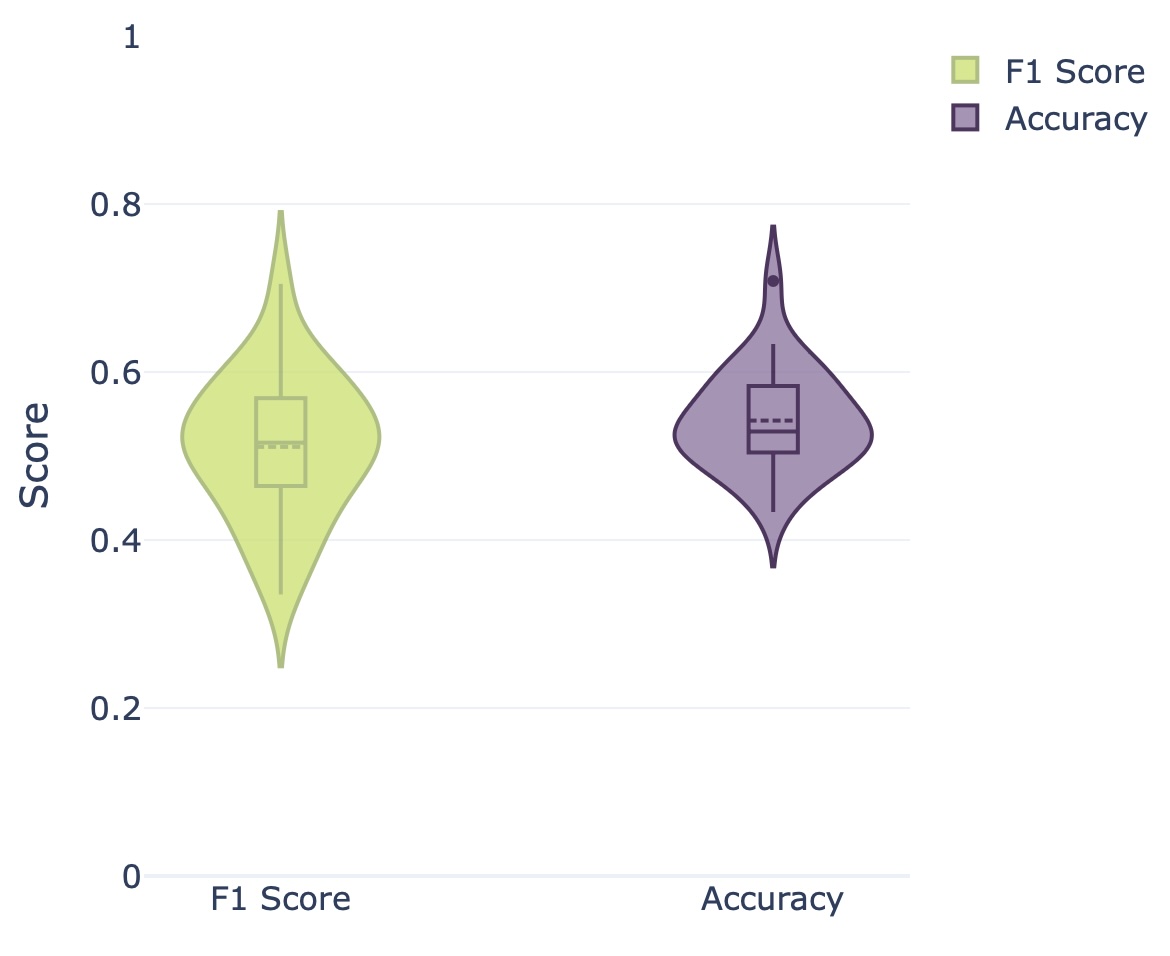}
  \caption{Distributions of per-subject F1 macro and accuracy for binary pain classification.}
  \label{fig:violin_plot}
\end{figure}

To visualize performance variability, Figure \ref{fig:violin_plot} shows the distributions of per-subject F1 macros and accuracies. The long tail suggests that while the model struggles overall, it may capture meaningful pain-related ECG patterns for specific individuals. This observation supports the potential of personalized or subject-adaptive modeling approaches.

%\begin{figure}[ht]
%  \centering
 % \includegraphics[width=1\columnwidth]{f1_acc_distribution.jpg}
  %\caption{Distribution of per-subject accuracy for binary pain classification.}
  %\label{fig:f1_accuracy_distribution}
%\end{figure}

\subsection{Representation space interpretability}

\begin{figure}[htbp]
    \centering
    \begin{subfigure}[b]{0.45\textwidth}
        \centering
        \includegraphics[width=\textwidth]{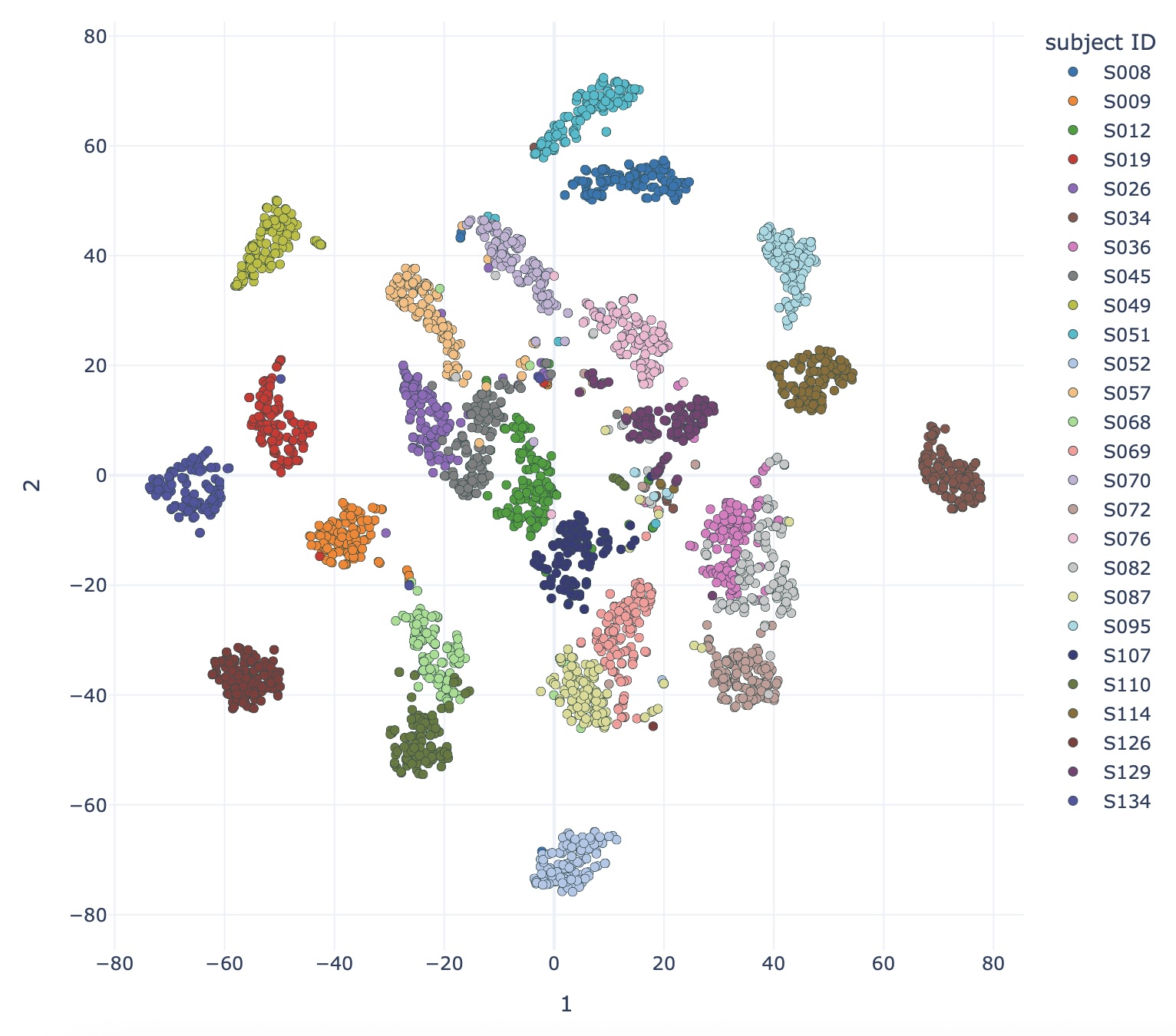}
        \caption{}
        \label{fig:tsne_subject}
    \end{subfigure}
    \hfill
    \begin{subfigure}[b]{0.45\textwidth}
        \centering
        \includegraphics[width=\textwidth]{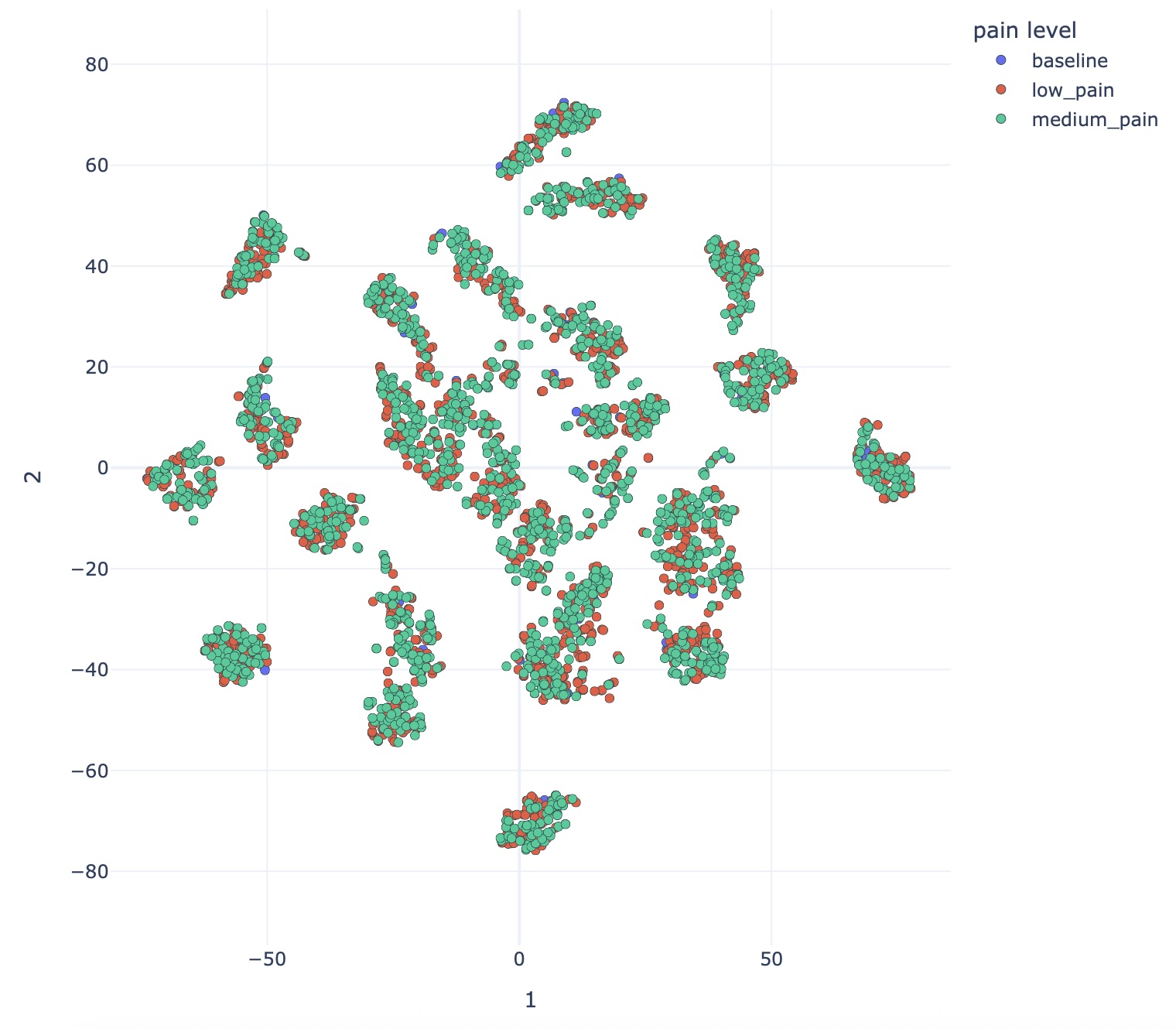}
        \caption{}
        \label{fig:tsne_pain}
    \end{subfigure}
    \caption{TSNE visualization of ECG representations (multimodal backbone) for 26 subjects from the training set, colored by (a) subject ID, (b) pain level.}
    \label{fig:tsne}
\end{figure}

In addition to downstream tasks such as classification and clustering, model performance can also be assessed through visual inspection of the learned representations. To enable such visualization, dimensionality reduction methods, such as TSNE \cite{van2008visualizing}, are typically applied to project high-dimensional embeddings into a lower-dimensional space. This allows for identifying structure, patterns, or groupings in the data, which can be examined in relation to the ground-truth labels. The TSNE visualizations of the training split of the X-ITE Pain dataset are presented in Fig. \ref{fig:tsne}

In Fig. \ref{fig:tsne_subject}, we observed that the representations often form subject-specific clusters, suggesting that the learned representations capture individual physiological characteristics. This subject-level separability indicates that personal traits significantly influence the structure of the data, which aligns with the known inter-individual variability in physiological responses. However, when examining the embeddings concerning pain levels (Fig. \ref{fig:tsne_pain}), no clear grouping or separation emerged. This suggests that pain-induced differences in the ECG signal are either subtle or masked by stronger individual patterns, making it challenging for the model to distinguish between pain states solely based on this modality.

\section{Discussion}

%ECG może zawierać mało informacji o bólu, albo inne składowe w ECG zaciemniają poziom bólu

%Task difficulty: Emphasize that binary pain classification from ECG alone remains a difficult problem, likely due to the weak signal-to-noise ratio and many confounders (e.g., stress, movement, baseline heart rate variability).

Our results suggest that pain recognition from ECG signals alone remains a challenging task. Despite using state-of-the-art self-supervised learning methods, the overall classification performance was limited. One possible explanation is that ECG may inherently contain little information directly related to pain, or that pain-related features are obscured by other dominant physiological components. Pain is often accompanied by changes in stress levels, breathing, and heart rate variability - factors that overlap and may act as confounders, making it difficult to isolate the specific signal components associated with pain.

%Inter-subject variability: Acknowledge that while average performance was low, some individuals had much higher classification results. This suggests that pain-related ECG patterns exist but may be subject-specific

We also observed a high degree of inter-subject variability in model performance. While the average classification accuracy was relatively low, some individuals exhibited notably better results. This indicates that pain-related patterns in ECG may be present but highly individual-specific. It reinforces the notion that personalized modeling or subject-adaptive techniques could be essential for improving recognition performance in this context.

%ACC w pretreningu poprawiło wyniki - dopisać może że to było n klatce piersiowej i mogło uchwycić informacje o unoszeniu się klatki piersiowej z ecg (przez typ pretreningu) 

Importantly, we found that incorporating accelerometer (ACC) data in the multimodal pretraining stage improved the downstream performance. Since the signals were collected from the chest, the ACC data likely captured subtle body movements associated with respiration. Through multimodal contrastive pretraining, the model may have learned to associate these respiratory cues with patterns in the ECG, thereby enhancing the representations derived from the ECG alone at inference time. This demonstrates the potential of cross-modal learning in uncovering latent dependencies between physiological signals that may otherwise remain underutilized in unimodal setups.

%\textcolor{red}{Model capabilities: Discuss whether SSL helped capture meaningful representations compared to using handcrafted features. If you didn’t compare directly, acknowledge that.?????}

\section{Limitations and Future Work}
%Limitations
%Unimodal input: ECG-only input restricts the physiological bandwidth. EDA or respiration may have captured pain better.

Despite these promising insights, several limitations must be considered to contextualize our findings and guide future improvements. First, the unimodal input using only ECG signals restricts the physiological information available for pain recognition. Other modalities, such as electrodermal activity (EDA) or respiration, might capture pain-related changes more effectively and improve model performance if incorporated.

%Generalization: Your model may not generalize well across subjects due to individual variability in ECG responses.

Second, the model's generalization across subjects remains limited due to substantial inter-subject variability in ECG responses. The unique physiological patterns of each person challenge the creation of universal pain detection models, underscoring the need for personalized or adaptive approaches.

%Label quality: Pain is subjective, and even high-quality datasets like X-ITE rely on self-report, which can introduce label noise.

Finally, the inherent subjectivity of pain introduces challenges related to label quality. Despite using a high-quality dataset such as X-ITE Pain, which relies on participant self-report, label noise and inconsistencies are inevitable. This uncertainty in ground-truth labels may hinder the model's ability to learn clear and consistent pain-related patterns.

%\textcolor{red}{Pretraining domain mismatch: Your SSL pretraining was done on real-world, non-pain data. The model may not have learned pain-specific representations.}

%Implications and Future Work

Recognizing these limitations opens important avenues for future research, highlighting potential strategies to enhance pain recognition models and better capture pain's complex, individualized nature.

%Toward personalization: Strong per-subject differences suggest potential in personalized or adaptive models that learn an individual baseline.
The strong inter-subject differences observed in this study suggest a promising direction toward personalized or adaptive models that establish individual baselines. Such approaches may better account for unique physiological patterns, improving pain recognition performance on a per-subject basis.

%Multimodal fusion: Suggest integrating EDA or other modalities if available in future work, even in low-cost wearable settings.
Future research should also explore multimodal fusion by integrating additional physiological signals such as EDA or respiration, especially when these modalities are accessible even in low-cost wearable devices. Leveraging complementary sources of information could help disambiguate pain-related changes from confounding factors.

%Improving SSL: Try task-specific pretraining or include pseudo-labels for weak supervision.

Improving self-supervised learning (SSL) techniques remains another key avenue. Task-specific pretraining or the inclusion of pseudo-labels for weak supervision may enhance the model’s ability to learn relevant pain-related features, leading to better downstream performance.

%Downstream task alternatives: Instead of binary pain classification, consider regression of pain intensity or pain trend detection over time.

Furthermore, when classifying multiple distinct pain levels in a non-binary setup, the ordinal relationship between classes can be lost. For example, low pain should logically be more similar to medium pain than to high pain. Future models should consider approaches that preserve these relationships, such as regression or custom loss functions, to reflect the pain intensity continuum better.

\section{Conclusions}
This study highlights both the potential and challenges of pain recognition using ECG signals collected in real-life conditions. While unimodal ECG-based models showed limited overall accuracy, the presence of substantial inter-subject variability suggests that personalized approaches may unlock more meaningful pain-related physiological patterns. 

However, the inherent complexity of pain as a subjective experience, combined with confounding physiological factors and label noise, underscores the difficulty of this task. Future research should focus on personalized modeling, multimodal fusion with signals such as EDA and respiration, and developing learning methods that better capture the ordinal and temporal nature of pain.

Ultimately, advancing pain recognition from wearable physiological signals requires a multifaceted approach, integrating methodological innovations with a nuanced understanding of individual differences and the underlying physiology.

\section{Ethical Impact Statement}

This work explores the feasibility of pain level assessment using electrocardiogram (ECG) signals and self-supervised learning to enable non-invasive and privacy-preserving physiological monitoring in real-world settings. ECG data can be collected passively and continuously via wearables, making it a less intrusive alternative to video, audio, or direct clinical observation. Importantly, our approach does not rely on sensitive personal content (e.g., facial expressions, speech) and focuses on a physiological signal already widely used in health monitoring.

However, pain is inherently subjective and context-dependent. Physiological signals may not always accurately reflect individual pain experiences, especially when influenced by emotional state, activity, or environmental conditions. Automated systems for pain detection carry the risk of false positives or false negatives, which could lead to inappropriate interventions or neglect. As such, any deployment of this technology should include human oversight and personalized calibration, and it should be used to support, not replace, clinical judgment.

We used de-identified data in compliance with ethical standards and made no attempts to infer sensitive attributes beyond those explicitly studied. Future research should further examine model fairness across age, gender, and health status groups to ensure that wearable-based pain monitoring does not exacerbate existing healthcare disparities.

\bibliography{bibliography}

\bibliographystyle{IEEEtran}

\end{document}